\documentclass[14pt,pdftex]{jetpl}
\usepackage[cp1251]{inputenc}
\usepackage[english]{babel}
\usepackage{graphicx}
\usepackage{amssymb}
\usepackage{amsmath}
\usepackage{amsfonts}

\twocolumn
\lat  
\title{Magnon spectrum in ferromagnets with a skyrmion}
\rtitle{Magnon spectrum in ferromagnets with a skyrmion}
\sodtitle{Magnon spectrum in ferromagnets with a skyrmion}
\author{D.\,N.\,Aristov$^{*+}$, S.\,S.\,Kravchenko$^{*+}$, and A.\,O.\,Sorokin$^{*+}$}
\rauthor{D.N. Aristov, S.S. Kravchenko, A.O. Sorokin}
\address{$^*$Department of Physics, Saint Petersburg State University, 199034, 7/9 Universitetskaya nab., St. Petersburg, Russia.}
\address{$^+$Petersburg Nuclear Physics Institute, NRC ``Kurchatov Institute'', 188300, Orlova Roscha, Gatchina, Russia}
\dates{\today}{*}

\abstract{
The analysis of the spin wave excitations in two-dimensional isotropic Heisenberg ferromagnet is performed with a single skyrmion in the ground state. We employ the ideas of semiclassical quantization method, duly modified for the use of the lattice model and Maleyev-Dyson boson representation of spin operators. The resulting Schr\"odinger equation for magnons describes the dispersion and wave functions of spin-wave excitations with strictly non-negative spectrum. In contrast to usual ferromagnet, we demonstrate the existence of three zero modes, corresponding to conformal symmetries spontaneously violated by the skyrmion configuration.
}

\begin{document}

\maketitle
\renewcommand{\refname}{}



\textbf{1.} Topological defects play an important role in condensed matter physics. The first and the most famous example is vortex lines, defining critical properties of type II superconductors in the external magnetic field \cite{Abrikosov1957}. In two dimensions, the role of defects is even more noticeable. So an interaction of vortices in the $O(2)$ model leads to emergence of a quasi-long-range order and a Berezinskii-Kosterlitz-Thouless (BKT) transition \cite{Berezinskii1971}. In $O(2)$ symmetric systems with the additional twofold degeneracy of the ground state, such as Josephson junction arrays in the magnetic field or XY helimagnets, vortex excitations with fractional charges lead to a phase transition on domain walls, and as a consequence to separation of a BKT and Ising (chiral) transitions \cite{Korshunov2002,Sorokin2012}. The appearance of so-called $\mathbb{Z}_2$-vortices corresponds to exceptional thermal properties of two-dimensional frustrated magnets with isotropic spins (see \cite{Kawamura2010} and Refs.\ therein). The superlattice structure observed in magnets \cite{Rossler2006, Muhlbauer2009, Yu2010, Nagaosa2013,Banerjee2014}
and multiferroics \cite{Seki2012} with the Dzyaloshinskii-Moria (DM) spin-orbit interaction in the magnetic field is believed to be related to vortex-like excitations, called skyrmions. The similar skyrmion structures appear in the quantum Hall systems \cite{Sondhi1993, Barrett1995, Ritz2013, Neubauer2009}.

In this paper we discuss topological defects in two-dimensional quantum ferromagnets. It is known that the usual $O(N)$ model, describing ferromagnets, has different types of topological defects. The case $N=1$ corresponds to the Ising model, where line-like defects are domain walls. The case $N=2$ has been mentioned above in a context of point-like vortices and a BKT transition. At $N=3$, defects of another type are present. They can be obtained as static classical solution of the $O(3)$ sigma model \cite{Belavin1975}, describing low-temperature properties of ferromagnets,
\begin{equation}
\label{HJ}
 H ={\cal A}  \int d^{2} \mathbf{r}\,\, \nabla_\mu\varphi^a\nabla_\mu\varphi^a, \quad \varphi\in S^2.
\end{equation}
Taking into account the isotropic condition at spatial infinity $\varphi(\infty)=\varphi_0$, the field $\varphi$ becomes a map $\varphi:\,\mathbb{R}^2\cup\{\infty\}\simeq S^2\to S^2$, which is characterized by an integer number $Q$, the topological degree of the map $\varphi$.

The families of solutions consist of configurations related to each other by global field-rotations and coordinate transformations. The latter symmetry includes rescalings (dilatations), that is specific to the two-dimensional sigma model, which is conformal invariant. As a consequence, a size of defects is not defined by the energy minimum conditions, in accordance with the Derrick theorem \cite{Derrick1964}. Nevertheless, a configuration with non-zero charge can not be continuously deformed to the true ground state, which is trivial ferromagnetic vacuum, $\varphi^a =\mathrm{const}$. Using rescalings, one can make the size of defect vanishing, but then a configuration becomes singular, and a charge $Q$ remains non-zero. In this sense, topological defects of the $O(3)$ model are stable. However, fixing the size is necessary to stabilize the defect in lattice formulations of the model. To fix the size of defects, the sigma model is routinely supplemented by high-order derivatives of the field (so-called Skyrme's terms) or other interaction terms \cite{Leese1991, Abanov1998}.

The energy $E_Q$ for topologically non-trivial configurations $Q\neq0$ is finite,  this is an important difference to vortices in the $O(2)$ model, where the vortex energy diverges logarithmically. The topological defects thus can be regarded as massive excitations of the $O(3)$ sigma model, in addition to usual massless (Goldstone) perturbative excitations. At non-zero temperature, topological defects are produced and destroyed by the thermal fluctuations, and the topological charge $Q$ is not conserved, but such  processes are suppressed by the factor $\exp(-E_Q/T)$. At zero temperature, the skyrmions become stable, as we mentioned above.

\begin{figure}[t]
\center{
\hspace{-5mm}
\includegraphics[width=0.68\linewidth]{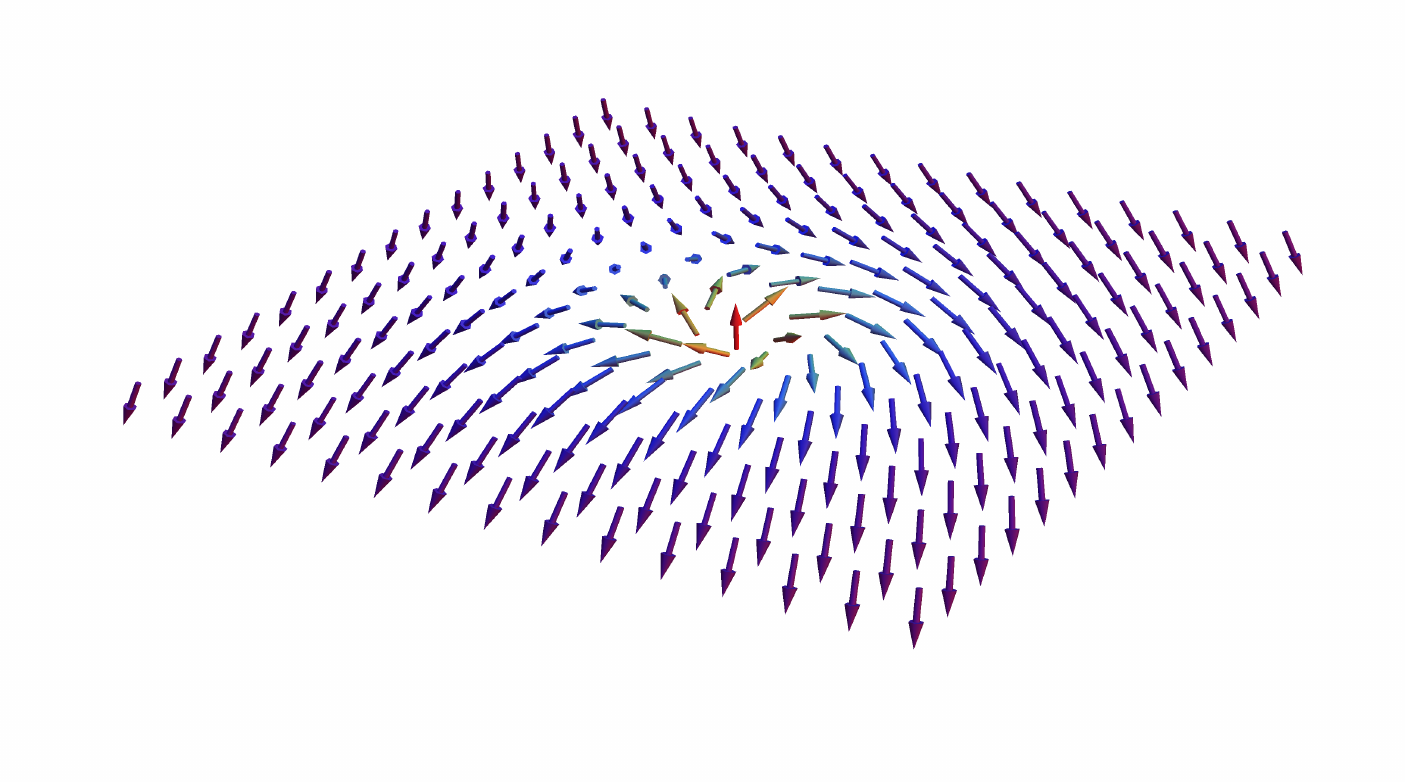}
\hspace{-5mm}
\includegraphics[width=0.38\linewidth]{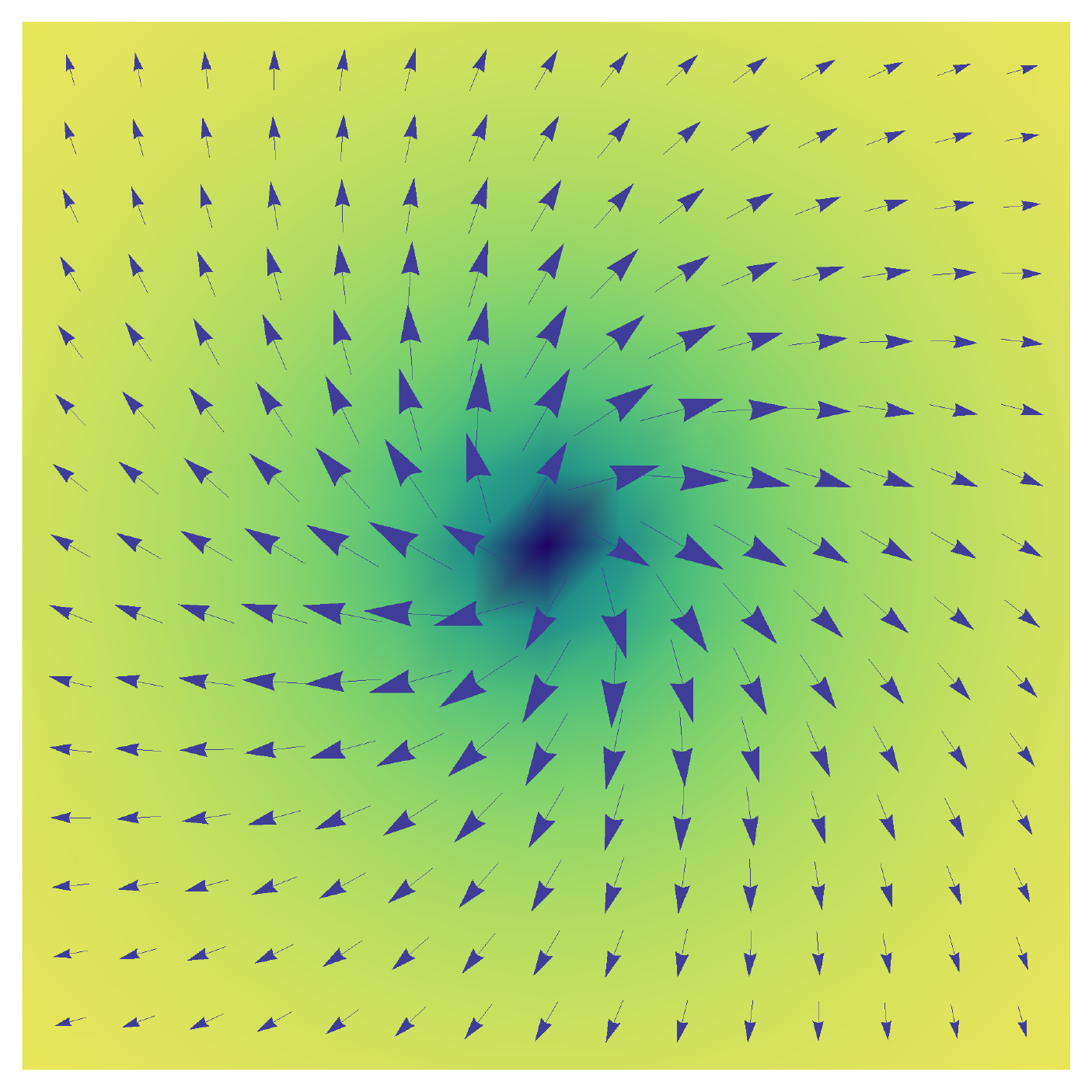}}
\caption{\label{fig1} Spin configuration in the radial parametrization of a skyrmion. A similar configuration is one block of a skyrmion lattice.}
\end{figure}

In this paper we consider the spectrum of spin-waves in the presence of a single skyrmion. Considerable theoretical efforts were devoted in recent years to the analysis of dynamics of magnetic fluctuations in magnets with skyrmions. The main motivation for these studies however is the case of a skyrmion lattice, which is stabilized by the DM-interaction and a sufficiently large applied magnetic field \cite{Han2010, Ezawa2010, Petrova2011, Mochizuki2012}. Strictly speaking, these additional interactions essentially change the symmetry of the Hamiltonian, and the single-skyrmion configuration ceases to be a local minimum of energy. Skyrmion-like configurations (fig.\ \ref{fig1}) appear as blocks of twisted modulated magnetic textures, which may be regarded either as of the topological origin \cite{Bogdanov94} or as a superposition of three helices \cite{Muhlbauer2009, Tatara14}. In either case, the scattering of magnons on a single skyrmion can be investigated \cite{Batista14, Nagaosa14, Garst14}.

\begin{figure}[t]
\center{
\hspace{-5mm}
\includegraphics[width=0.68\linewidth]{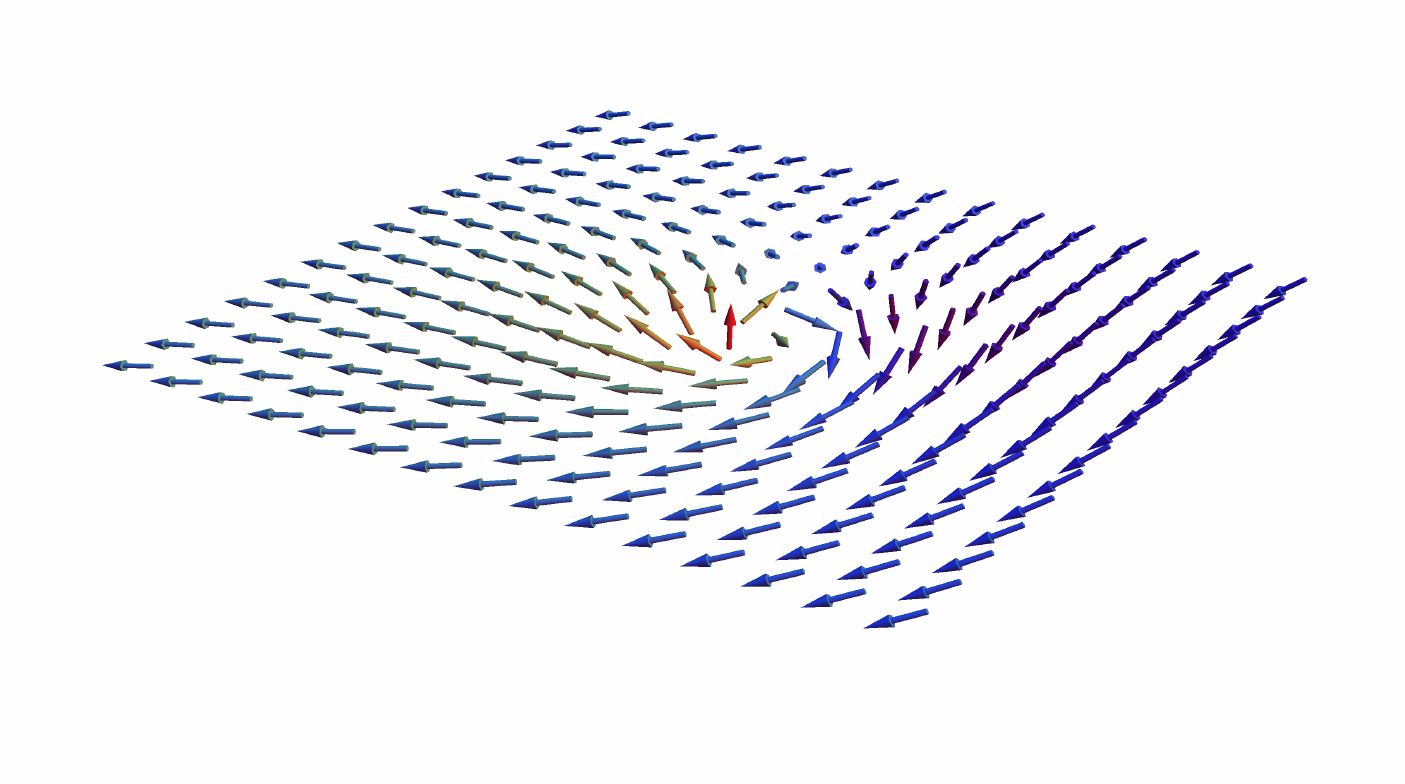}
\hspace{-5mm}
\includegraphics[width=0.38\linewidth]{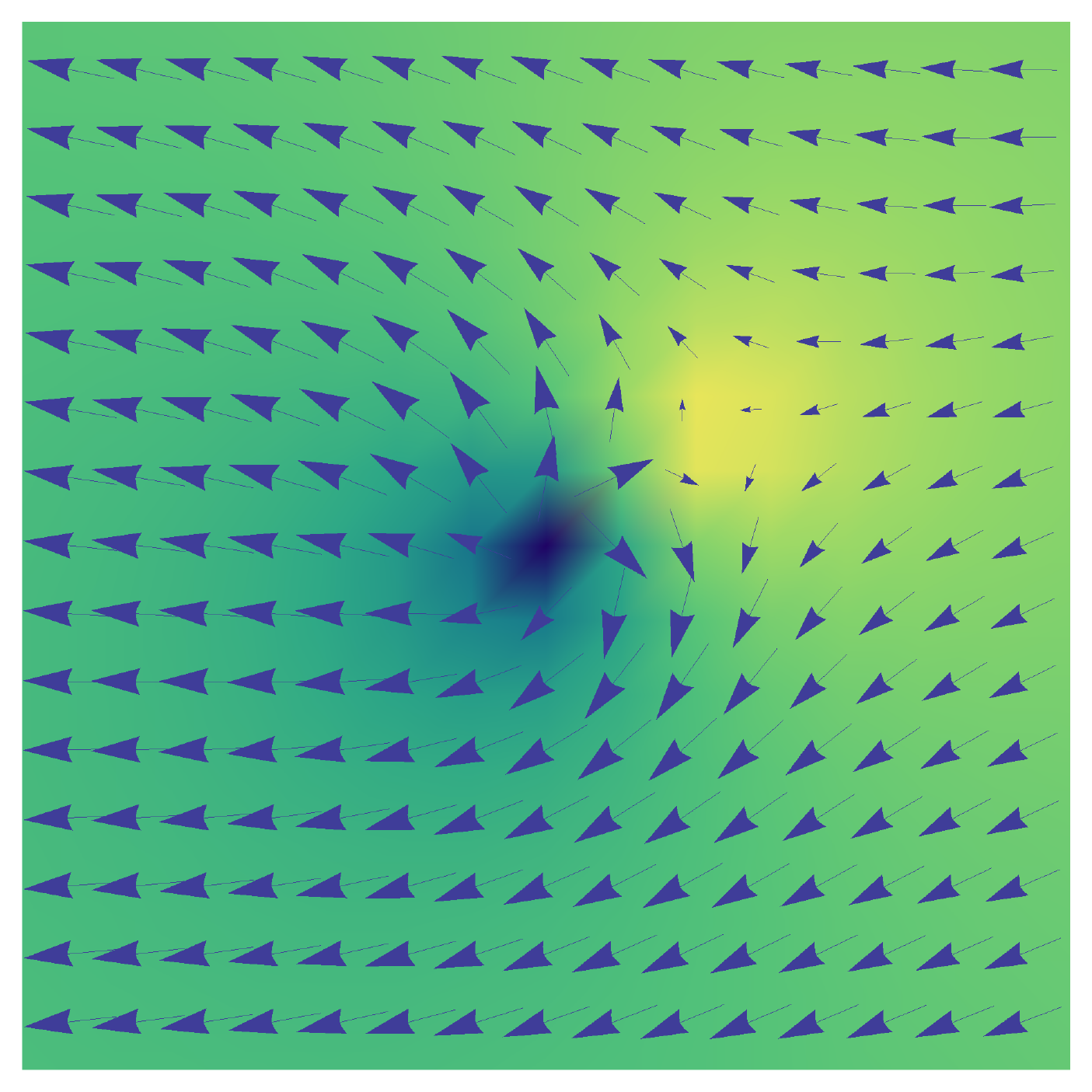}}
\caption{ \label{fig2} Spin configuration of the skyrmion presented in Fig.\  \ref{fig1} after a special conformal transformation of spatial coordinates or, equivalently, after a global spin rotation followed by certain translation. The configuration appears like a vortex-antivortex pair, when judged by only one spin projection.}
\end{figure}

Similar modulated textures containing skyrmion-like blocks have also been observed  in other systems, e.g.\ 1.5-type superconductors \cite{Babaev14}. Such systems have stabilizing interactions of different symmetry properties, and consequently a different spectrum of excitations. In the absence of lower-symmetry interactions, a parametrization of the skyrmion solution can be chosen in an unusual way (see Fig.\ \ref{fig2}). It shows essential differences between the {\it bona fide} Belavin-Polyakov skyrmion and a vortex-like object in magnets with the DM-interaction.

The problem of quantum excitation spectra in topologically non-trivial background is rather general one, and apparently has its peculiarities for each type of topologically protected ground state. It is known several approaches to this problem, including non-perturbative ones \cite{Rajaraman1982}. We employ the duly modified method of semiclassical quantization, based on the lattice Heisenberg model and the Maleyev-Dyson representation of spin operators. This method is natural for spin systems. It is routinely used for plain ferromagnetic state, provides us with both an equation for a spectrum and magnon-magnon interaction terms, and allows to compare spectra in trivial and non-trivial backgrounds. Although, this method can be generalized to systems with other types of topological defects and/or additional spin interactions, we investigate the pure Heisenberg model as a instructive case with the fully analytic character of intermediate formulas and solutions.

\textbf{2. }
Consider the exchange lattice Hamiltonian
\begin{equation}\label{H}
H =     \sum_{i,j} J(\mathbf{r}_i - \mathbf{r}_j)  \mathbf{S}_{\mathbf{r}_i} \mathbf{S}_{\mathbf{r}_j}
\end{equation}
We assume that the ground state is characterized by the non-collinear skyrmion ground state.
Our aim is to rewrite the Hamiltonian  (\ref{H}) in such local basis, where the average local spin is directed along the $\hat z$-axis. The transition to this local basis, $\mathbf{S}_{\mathbf{r}} = \widehat U(\mathbf{r})\tilde{\mathbf{S}}_{\mathbf{r} } $, is given by the position-dependent 3$\times$3 matrix, 
 $\widehat U(\mathbf{r}) =  e^{-\alpha  \sigma_3}e^{-\beta  \sigma_2}e^{-\gamma  \sigma_3}$, 
with $\sigma_3$, $\sigma_2$ generators of SO(3) group, and  $\alpha$, $\beta$, $\gamma$ Euler angles.
In the new basis the Hamiltonian (\ref{H}) takes the form:
\begin{equation}
\label{H1}
H =    \sum\nolimits_{\mathbf{r}  , \mathbf{n} }J(\mathbf{n})  \tilde{\mathbf{S}}_{\mathbf{r} } \widehat R\left(\mathbf{r} , \mathbf{n}\right)\tilde{\mathbf{S}}_{\mathbf{r} +\mathbf{n}  }
\end{equation} with 
$\mathbf{n} = \mathbf{r}_i - \mathbf{r}_j$, $\widehat R\left(\mathbf{r} , \mathbf{n}\right) = \widehat U^{-1}(\mathbf{r} ) \widehat U(\mathbf{r} +\mathbf{n})$.

Assuming $J(\mathbf{n})$ rapidly decreasing with distance $\mathbf{n}$, we expand the matrix  $R^{ab}(\mathbf{r},\mathbf{n})$ in a series  :
\begin{equation}\label{R}
R^{ab}(\mathbf{r},\mathbf{n}) =\delta_{ab}+ \chi^{ab}_{1,\mu}(\mathbf{r})n^\mu + \tfrac12 \chi^{ab}_{2,\mu \nu}(\mathbf{r})n^\mu n^\nu + \ldots
\end{equation}
with  $\delta_{ab}$ Kronecker symbol and
\begin{equation}
\label{def:chi12}
 \begin{aligned}
 \chi^{ab}_{1,\mu}(\mathbf{r}) &= U^{c a}(\mathbf{r})\nabla^\mu  U^{c b}(\mathbf{r})  \,, \\
 \chi^{ab}_{2,\mu \nu}(\mathbf{r}) &= U^{ca}(\mathbf{r})\nabla^\mu \nabla^\nu U^{cb}(\mathbf{r})  \,.
 \end{aligned} \end{equation}
Here  and below we assume the summation over the repeated tensorial indices.

The equilibrium state of the spin configuration implies that the total field induced by the neighboring spins is parallel to the direction of the spin at a given site, i.e.\ along the direction $\hat e_{3}$. This results in a double condition

\begin{equation}
\label{equil}
 \sum_{\mathbf{n}}J(\mathbf{n})R^{a3}(\mathbf{r},\mathbf{n})= 0 \,, \quad
 \sum_{\mathbf{n}}J(\mathbf{n})R^{3a}(\mathbf{r},\mathbf{n})= 0 \,,
 \end{equation}
with  $a = 1,2$.

The explicit dependence of  $U$, $\chi^{ab}_{1,\mu}$, $\chi^{ab}_{2,\mu \nu}$ on the Euler angles is known.
The conditions   (\ref{equil}) then determine the  dependence of these angles on $\mathbf{r}$.
We  assume here that the average spin has the same absolute value, which is verified below.

Eq.\  (\ref{equil}) does not pose any restriction on $\chi^{ab}_{1,\mu}$, in centrosymmetric situation, when   $J(\mathbf{n})=J(-\mathbf{n})$   and $\sum_{\mathbf{n}}J(\mathbf{n})\mathbf{n} = 0$.  
Putting $\tilde{ {S}}^{b}_{\mathbf{r} } =  s \, \delta_{3b}  $ we obtain from  the first equilibrium condition 
\footnote{It can be shown, that isotropic quartic terms in the Taylor expansion \eqref{R}  result in the modification 
of Eq.\  \eqref{skyr} , $  \beta = 2 \arctan(r_0/r) +  \beta_1(r/r_0) {\cal O}((a_{0}/ r_{0})^{2}) $ 
with $a_{0}$ lattice constant and 
$ \beta_1(\rho) = 
 { \rho  \left(\left(\rho ^2+4\right) \rho ^2+2 \left(\rho
   ^2+1\right) \log \left(\rho ^2+1\right)\right)}/{ \left(\rho
   ^2+1\right)^2} ,$   
whereas the skyrmion energy $4\pi Cs^{2}$ may acquire relative corrections of order $ (a_{0}/ r_{0})^{4}$, which we neglect. 
   }
   in (\ref{equil}) :
\begin{equation}
\label{chik3}
\sum_{\mu} \chi^{a3}_{2,\mu\mu}(\mathbf{r}) = 0\, ,\quad a = 1,2
\end{equation}
which might be represented as one equation for complex-valued quantity 
\begin{equation} \label{chi2+}
\begin{aligned}
\chi^{+}_{2} & \equiv \chi^{13}_{2,\mu\mu} + i \chi^{23}_{2,\mu\mu}  \\ 
  &= ie^{-i\gamma} \left[2 \cos\beta \nabla \alpha \nabla \beta + \sin \beta\, \nabla^{2} \alpha
\right. \\ & \left.
- i (\nabla^{2} \beta - \sin \beta \cos \beta (\nabla \alpha)^2 ) \right] =0 \,.
 \end{aligned} \end{equation}
The real and imaginary parts of the expression in square brackets here are proportional to the variational derivatives of
the classical energy over $\alpha$ and $\beta$, respectively.

The equation $\chi^{+}_{2} = 0$ supplied by the conditions
\begin{equation} \label{skyr}
\frac{d \beta}{d \phi}=0   ,\quad
\frac{d\alpha}{d\phi} =   \ell , \quad   \frac{d\alpha}{dr} = 0
\end{equation}
with $\ell =1 $ for a single skyrmion centered at the origin,  $\mathbf{r} = 0$, leads to the solution
\begin{equation}\label{beta}
\beta = 2 \arctan\left({r_{0}}/{r}\right)\,, \quad  \alpha + \alpha_0 = \phi
\end{equation} with
$r_0$ Skyrmion radius. It translates to the well-known explicit dependence of the local spin direction on the coordinates
\begin{equation}
\label{skyr-expl}
\begin{aligned}
S^{1} + i S^{2} &= s\frac{2r r_0}{r^2 + r^2_0} e^{ i\left(\phi- \alpha_0 \right)} \,, \quad
S^{3} = s\frac{r^2 - r^2_0}{r^2 + r^2_0}  \,.
 \end{aligned} \end{equation}
Notice that  $r_0$ and the phase $\alpha_0$  are not  determined in this calculation.

The Euler angle $\gamma$  is arbitrary and not defined from (\ref{chi2+}). The  rotation by $\gamma$ gives a transformation $ \tilde{S}^{\pm }_{j} \to  \tilde{S}^{\pm}_{j} e^{\pm i\gamma}$, which reduces to $a_{j} \to a_{j} e^{i\gamma}$ in terms of bosons below. The natural choice, $\widehat U = 1$ at  $r\to\infty$, would correspond to $\gamma = - \alpha  = \alpha_{0} - \phi$, one can check that it results in discontinuity of  $\widehat U$ at  $r\to0$. Alternatively, demanding the continuous character of $\widehat U$ at the origin, $r=0$, means that $\alpha -\gamma = cst$ and it translates to the double full rotation, $\widehat U = \exp[ 2(\phi -\alpha_{0}) \sigma _{3}]$ at   $r\to\infty$. We adopt below the latter choice,
\begin{equation}\label{gamma}
\gamma = \alpha = \phi -\alpha_{0} \,,
\end{equation} which provides us with the continuity at $r=0$.


\textbf{3.}  Knowing the  form of   $\widehat U$ via the $\mathbf{r}$-dependent $\alpha$, $\beta$, $\gamma$ given by (\ref{skyr}) and  (\ref{beta}), we obtain the explicit expressions  $\chi^{ab}_{1,\mu}(\mathbf{r})$ and  $\chi^{ab}_{2,\mu\mu}(\mathbf{r})$. We then use these expressions  and
 Maleyev-Dyson representation for spin operators, preserving the spin commutation relations,  $[\tilde{S}^a,\tilde{S}^b] = i \epsilon_{abc}\tilde{S}^c$:
\begin{equation} \begin{aligned} \label{boz}
     \tilde{S}^{3}_{j} &=s-a^+_{ j} a_{ j} \,, \quad
       \tilde{S}^{+}_{j}=\sqrt{2s}a_{ j}  \\
     \tilde{S}^{-}_{j} &=\sqrt{2s}\left( a^{\dagger}_{ j} - \tfrac{1}{2s}a^\dagger_{ j}a^{\dagger}_{ j}a_{ j} \right)
  \end{aligned}  \end{equation}
here $s$ the value of spin,  $\tilde S^{\pm} _{j}= \tilde S^{1}_{j} \pm i \tilde S^{2}_{j}$ and $[a_{ j},a^+_{ j}] = 1$.  We thus express our Hamiltonian  (\ref{H1})  in bosonic representation.

We make two simplifications now. First is the long wavelength limit of our model, corresponding to \eqref{HJ} : 
\begin{equation}\label{J}
J(\mathbf{q}) = \sum_{\mathbf{n}}e^{i\mathbf{q}\mathbf{n}} J(\mathbf{n})  \simeq J(0) + \tfrac{1}{2}Cq^2
\end{equation}
with $J(0) < 0$ for ferromagnetic exchange and $C > 0$. Another simplification  is the semiclassical limit of large spin $s$. In this sense, Eq.\ (\ref{HJ}) corresponds to the limit $s\to \infty$. Assuming $s\gg 1$, we keep those largest-in-$s$ terms, which will provide us with  the spectrum of magnon excitations.   We write to this end
\begin{equation}
\label{linearS} \begin{aligned}
 \tilde{S}_{\mathbf{q}}^{1} & 
 \simeq  \frac{\sqrt{2s}}{2} \sum_{ j} e^{i\mathbf{q}\mathbf{r}_j}(a_{ j}^{\dagger} + a_{ j})
 = \frac{\sqrt{2s}}{2}(a_{-\mathbf{q}}^{\dagger} + a_{\mathbf{q}})
\\
 \tilde{S}_{\mathbf{q}}^{2} & 
 \simeq \frac{\sqrt{2s}}{2i} \sum_{ j} e^{i\mathbf{q}\mathbf{r}_j}(a_{ j} - a_{ j}^{\dagger}) = \frac{\sqrt{2s}}{2i}(a_{\mathbf{q}} - a_{-\mathbf{q}}^{\dagger})
\\
 \tilde{S}_{\mathbf{q}}^{3} & = s \delta(\mathbf{q}) - \sum_{\mathbf{k}}a^\dagger _{\mathbf{q}+\mathbf{k}}a_{\mathbf{k}}
  \end{aligned}  \end{equation}
Putting these expressions into the Hamiltonian we obtain the classical energy of the magnet in the order $s^{2}$ of the form
\begin{equation}
s^{2} \int d\mathbf{r}\, \left(  - J(0)  +  C \frac{  4 r_0^2}{{(r^2 + r_0^2)^2}} \right),
\end{equation}
here the first term gives the energy of uniformly magnetized sample, and the second contribution, $\delta E = 4 \pi C s^{2} > 0$ shows that the skyrmion configuration is higher in energy and independent of its size $r_{0}$.

The terms linear in bosons possess the prefactor $s^{3/2}$ and vanish, due to the  condition on $\chi^{a3}_{2,\mu\mu}$, Eq.\  (\ref{chi2+}). (The terms $\chi^{3a}_{2,\mu\mu}$ are non-zero and  lead to linear-in-bosons contribution which is exactly compensated by contribution from the $\chi^{ab}_{1,\mu}$ after integration by parts.)

The terms linear-in- $s$ are quadratic in bosons, stem from both $\chi^{ab}_{1,\mu}$ and $\chi^{ab}_{2,\mu\mu}$ and have the form:

\begin{equation}
\label{Hquad}
\begin{aligned}
H &=Cs \int d\mathbf{r}\,  a^\dagger_{\mathbf{r}} \, \widehat {\cal H} \, a_{\mathbf{r}}  \,, \\
\widehat {\cal H} & = -     \nabla^2  +  \frac{4  }{ r^2 + r_0^2}  L_z  +
 4 \frac{ r^{2} -  r_0^2}{{(r^2 + r_0^2)^2}}      \,,
\\
\nabla^{2} & =  \frac1r \frac{\partial }{\partial r} r \frac{\partial }{\partial r} - \frac{L_z^{2}}{r^2} , \quad
L_z = -i \frac{\partial}{\partial \phi} \,.
\end{aligned}
\end{equation}
In consistency with our approximation (\ref{linearS}) we neglect  the terms  of order of $s^{1/2}$, $s^{0}$, which contain cubic and quartic boson combinations, respectively, and correspond to interaction of magnons.

The Hamiltonian  (\ref{Hquad}) assumes the above choice $\gamma=\phi$, corresponding to continuity of $\widehat U$ at $r=0$.  The usual magnon spectrum of ferromagnet is described by the first term, $\nabla ^{2}$, in  $\cal H$ and the second and the third terms describe the magnons in the presence of the skyrmion. The second term  appears  due to the chiral character of the skyrmion and should be viewed as a scalar product $L_{z} \ell$, so that for the antiskyrmion with $\ell = -1$ this term changes its sign. Formally we restore the usual ferromagnet case by putting $r_{0}\to \infty$; alternatively, we may set $r_{0}=0$ and shift $L_{z}\to L_{z}-2 $, which corresponds to $\gamma = -\alpha$.

Next, we see the absence of the anomalous terms of the form $a_{\mathbf{r}}a_{\mathbf{r}}$, $a^{\dagger}_{\mathbf{r}}a^{\dagger}_{\mathbf{r}}$  in  (\ref{Hquad}), which appears in antiferromagnets and in case of a skyrmion with stabilizing interactions \cite{Garst14}. As a result, we have the absence of zero-motion effects, and the average spin has its saturated value, $s$, at zero temperature.


Applying the equation of motion, $i  \frac{\partial}{\partial t} a_{\mathbf{r}} = \left[ a_{\mathbf{r}} ,H\right]$ and using the separation of variables,  we express our operators via eigenmodes:
$ 
a_{\mathbf{r}} (t) = \sum_{E,m} e^{-i E t} e^{i m\phi}  \psi_{m,E}(r) a_{m,E}
$ 
where $[a_{m,E},a^{\dagger}_{m',E'}]= \delta_{mm'}\delta(E-E')$ and $a^{\dagger}_{m,E}$ creates the magnon with angular momentum $m$ and energy $E$. Given the absence of translational and mirror symmetry, the momentum is not a good quantum number and our equations are not invariant upon the change $m\to -m$.

The equation for the radial part of eigenfunctions $\psi_{-m,E}(r)$ is obtained from \eqref{Hquad}
by putting $L_{z} = -m$ 
and can be solved numerically.
 Substituting 
  $z = 1+2 (r/r_{0})^{2}$,  $\varepsilon = E r^{2}_{0}/ Cs$, $\psi_{-m,E}(r)  = (z-1)^{-1/2} \Upsilon(z)$ we reduce the  Schr\"odinger equation $\widehat{\cal H}\psi = E \psi$  to  the form 
 \[  \frac{d^{2} }{d z^{2}}     \Upsilon  = \left[
-\frac{m^2-1}{4 (z-1)^2}-\frac{2}{(z+1)^2}  +\frac{1-m}{z^2-1}+\frac{\epsilon/8 }{
   (z-1)}  \right]  \Upsilon 
\]

This equation has two regular ($z=\pm1$) and one essential ($z=\infty$) singular points, it resembles the equation for Coulomb spheroidal functions \cite{Komarov1976}, but is more complicated. It 
is not of hypergeometric type and its solution in any basis of hypergeometric functions is given by infinite series, which is truncated only in special cases. Below we show that  compact analytical form of the solution is found   for either $\epsilon =0$ or $m=0$.

\textbf{4.} 
The direct substitution shows that there is a whole series of eigenfunctions $\psi$ with  $E = 0$ and different $m$.
\begin{equation}
\label{zmodes}
\psi_{-m,0} \propto  \frac{r^{m}}{r_0^2 + r^2}  \,.
\end{equation}
Most of these solutions are divergent either at $r=0$ or at $r\to\infty$. The exceptions are   $\psi_{-m,0} $ with $m=0,1,2$. For other values of $m$ the effective potential in \eqref{Hquad} is strictly positive which means non-negative eigenvalues. Numerics also show that all eigenvalues of \eqref{Hquad} are non-negative, which means \emph{stability} of the skyrmion solution in the absence of stabilizing interactions.  This should be contrasted with negative eigenvalues for the quantum modes \cite{Garst14}, indicating instability of a single skyrmion at certain values of stabilizing interactions.

One can associate zero modes with global continuous transformations of the Hamiltonian, which change the field configuration but not the energy, according with the Goldstone theorem.  These transformations should manifest themselves as static field fluctuations, and they are not well described in terms of second-quantized bosons but rather  in first-quantization scheme.

Our field configuration is characterized by $\widehat U_{0}(\mathbf{r})$ with the skyrmion described by Eqs.\  (\ref{beta}), (\ref{gamma})  and centered at $\mathbf{r}=0$.  After infinitesimal change of coordinates  $\mathbf{r} \to \mathbf{r} +  \mathbf{r}_{1}$ we write
 \begin{equation}
\widehat U_{0}(\mathbf{r} +\mathbf{r}_{1}) \simeq  \widehat U_{0}  (\mathbf{r})
\left (1 +  {r}_{1} ^{\mu}\, \widehat \chi_{1,\mu}(\mathbf{r}) \right)
\,.
\end{equation} see (\ref{def:chi12}).
In the classical limit, $s\to \infty$, Eqs. (\ref{boz}) become
\begin{equation} \begin{aligned} \label{classicalS}
     \tilde{S}^{3} &=s- \xi^{2} - \eta^{2} \,, \quad
       \tilde{S}^{\pm} = \sqrt{2s}  (\xi \pm i \eta),
\end{aligned}  \end{equation}
and the next term in $1/s$ is
\begin{equation}
\tilde S^{\pm} = s\, {r}_{1} ^{\mu} \left(\chi^{13}_{1,\mu}(\mathbf{r}) \pm i    \chi^{23}_{1,\mu}(\mathbf{r})
\right) \,.
\end{equation}
Letting $\mathbf{r}_{1} = r_{1} (\cos\phi_{1}, \sin\phi_{1})$,  we obtain 
\begin{equation}
\label{shift}
\tilde S^{\pm} = -2is\, {r}_{1} \frac{e^{\mp i(\phi_{1}-\alpha_{0})}}{1+(r/r_{0})^{2}}  \,.
\end{equation}
In this terms, the infinitesimal form of translations, dilations (plus rotations) and special conformal transformations is
\begin{equation}
\label{trans-inf-r1}
\begin{aligned}
r_{1} & = b , \quad \phi_{1} = cst\,, \\
r_{1} & = b r , \quad \phi_{1} = \phi \,, \\
r_{1} & = b r^{2} , \quad \phi_{1} = 2\phi + cst \,.
\end{aligned}
\end{equation}
It becomes now evident that Eqs.\ (\ref{shift}), (\ref{trans-inf-r1}) correspond to (\ref{zmodes}) with $m=0,1,2$.


\textbf{5. }
Knowing zero modes (\ref{zmodes}), we can transform the equation $\widehat{\cal H}\psi = E \psi$  according to the recipe of supersymmetric (SUSY) quantum mechanics. \cite{Cooper1995} In doing so, we hope to arrive at a simpler form of the potential term which will allow us to find eigenfunctions exactly.  
Introducing $x=r/r_{0} $ and  $ \Phi(x)= x^{1/2} \psi (x r_{0})$ we reduce our equation  to the form
$   A ^{\dagger}A \Phi  = \varepsilon \Phi  $ 
with 
\begin{equation}
\begin{aligned}
A & =   \frac{d  }{d x}  + W(x) \,,\quad
A ^{\dagger}=  -  \frac{d  }{d x }  + W(x) \,, \\
 W(x) & =  - \frac{d \ln \Phi_{0}}{d x } = - \frac{m+1/2}{x} + \frac{2 x }{1 + x^2}
\end{aligned}
\end{equation}
and $\Phi_{0}= \frac{x^{m+1/2}}{1 + x^2} $ zero mode. The SUSY partner equation is given by

\begin{equation}
\begin{aligned}
A A ^{\dagger} \Phi & = \left (-  \frac{d^{2} }{d x^{2}}  +V_{2}(x)  \right)\Phi
= \varepsilon \Phi  \\
V_{2}(x) 
& =
 \frac{(m+1)^{2}-1/4}{x^2} -  \frac{4m}{x^2 +1}
\end{aligned}
\label{SUSYpartner}
\end{equation}
We notice that the partner equation  at $m=0$ corresponds to the free motion with shifted angular momentum and its solution $ \tilde \Phi_{\varepsilon}$ is readily found. The solution to the original equation with the same energy is given by
$\Phi _{\varepsilon} = A ^{\dagger} \tilde \Phi_{\varepsilon}$.
After simple algebra we  obtain
\begin{equation}
\begin{aligned}
 \psi_{m=0,E} (r) 
 & =\frac { r_{0}^{2} J_{0} ( \kappa r) - r^{2}  J_{2} (\kappa r)  }
 {r_{0}^{2}+r^{2}} \,.
\end{aligned}
\label{exact-m0}
\end{equation}
with the analog of wave-vector, $\kappa = \sqrt{E/Cs}$.
It is seen here, that the exact wave function smoothly  interpolates between $J_{0} (r\kappa)$  (i.e.\ free motion with $m=0$) at $r\ll r_{0}$  and  $J_{2} (r\kappa) $ (free motion with $m=2$) at $r\gg r_{0}$. Interestingly, the differential equation equivalent to our Hamiltonian (\ref{Hquad}) (with $E \sim C s \kappa$ instead of our $E = C s \kappa^{2}$) and the exact wave function of the form (\ref{exact-m0}) were obtained in \cite{Belov1997} in the analysis of magnon dispersion for $s=1/2$ antiferromagnet.


\textbf{6.}
Summarizing the results, we note that the method presented in this paper for the analysis of spin-wave excitations in the topologically non-trivial background includes a combination of two well known steps. The first one is a suitable rotation of local frames for spin operators, which maps non-trivial configuration as a local minimum of classical energy. The second step is the use of Maleyev-Dyson representation for spin operators, which is a semiclassical method (large spin assumption) in its essence. For the pure Heisenberg model with a skyrmion, we obtain the Schr\"odinger equation in the linear spin-wave approximation in analytic form, and find its explicit ``s-wave'' solution. For the standard ferromagnetic uniform ground state, the magnons correspond to usual plane waves, and the zero (Goldstone) mode wave function tends unity $e^{i\mathbf{qr}} \to 1$.  By contrast, the skyrmion ground state configuration is characterized by internal degrees of freedom, which are translations, rotation and dilatation, in addition to the spontaneous direction of magnetization at the infinity. As a result, one sees three zero modes in the equation for magnons, instead of one. This is quite unusual and we trace these modes to the conformal symmetries of our Hamiltonian. We proved that the energies are strictly non-negative, the influence of the the skyrmion amounts to peculiar potential term, which vanishes in the limits of small and large skyrmion radius.

Our construction allows to take into account the terms of magnon interaction naturally in comparison to some other approaches. One expects only the appearance of a finite number of magnon vertices, with three, four and five magnons interacting. Therefore using the Maleyev-Dyson representation is preferred over the Holstein-Primakoff representation, the latter leading to infinite number of interaction terms. The corresponding analysis \cite{future} shows that the interaction of magnon is more pronounced in the presence of the skyrmion than in case of plain ferromagnetic, but only of marginal importance, and the magnons remain well-defined excitations.


We thank U.K.  R\"o\ss{}ler, S.V. Grigoriev, S.V. Maleyev, J. Schmalian and M. Garst for useful discussions.
We acknowledge Saint-Petersburg State University for research grants 11.38.636.2013 and 11.50.2514.2013.

\end{document}